# Estimations of First $2^+$ Energy States of Even-even Nuclei by Using Artificial Neural Networks

Serkan Akkoyun[a,b,*], Hüseyin Kaya[a], Yunis Torun[b,c]

[a]*Department of Physics, Sivas Cumhuriyet University, Sivas, 58140, Turkey*

[b]*Artificial Intelligence Systems and Data Science Application and Research Center, Sivas Cumhuriyet University, Sivas, 58140, Turkey*

[c]*Department of Electric and Electronic Engineering, Sivas Cumhuriyet University, Sivas, 58140, Turkey*

**Abstract**

The first excited $2^+$ energy states of nuclei give much substantial information related to the nuclear structure. Including these levels, all excited states of nuclei are shown regularities in spin, parity and energy. In the even-even nuclei, the first excited state is generally $2^+$ and the energy values of them increase as the closed shells are approached. The excited levels in nuclei can be investigated by using theoretical nuclear models such as nuclear shell model. In the present study for the first time, we have used artificial neural networks for the determination of the energies of the first $2^+$ states in the even-even nuclei in the nuclidic chart as a function of Z and N numbers. According to the results, the method is convenient for this goal and one can confidently use the method for the prediction of the first $2^+$ state energy values whose experimental values do not exist in the literature.

**Keywords:** First $2^+$ energy, artificial neural network, even-even nuclei

PACS: 21.10.-k, 21.60.-n, 84.35.+i

Corresponding author e-mail: sakkoyun@cumhuriyet.edu.tr; Phone: +90 346 219 1010

# 1. INTRODUCTION

By using a central attractive force, it is found that the ground state is $0^+$ and the first excited state is generally $2^+$ of even-even nuclei [1]. The energy value of the first excited state in nuclei depends regularly on the proton and neutron numbers [2]. Whereas, some regularities are observed in the energy values of the first excited state and these regularities can be explained in terms of mixtures of states of excitation of different neutron and proton pairs [3]. Both the jj coupling model and the liquid drop model of the nucleus can explain the regularities in spins and parities of first and second excited states of even-even nuclei [4]. In even-even nuclei, the first excited state includes much information about the nuclear structure such as deformation and shape of nuclei, lifetime of the nuclear states and transitions between levels. Some information about the proton and neutron interactions in partially-filled shells might also be obtained from the first excited state. The energy values of these states increase as the closed-shell is approached. Namely, closing a shell causes a sharp increase in the energy of the first excited state [5]. Therefore, first excited state energy and spin values are sensitive to the shell structure of the nuclei [6]. According to the shell model (SM) of the nucleus, large shell gaps are observed between the shells for stable nuclei. These nuclei have magic proton and neutron numbers whose values are 2, 8, 20, 28, 50, 82 and 126 [7]. Due to these large gaps, large transition energy values are measured between the first $2^+$ and ground states in nuclei. Thus, the transition probability from the first excited state to the ground-state decreases monotonically. Related to these levels, the second excited state has about twice as much energy [8] and if we excite an odd A nucleus, we would expect to find its first excited state at least as low as that of its even-even core.

Theoretically, the predominance of spin 2 and even parity first excited states of even-even nuclei has been explained by using nuclear SM. The first excited state of nuclei is assumed due to the excitation of a single pair of nucleons. If the proton (neutron) shell is closed in nuclei, the first excited state is ascribed to neutron (proton) excitations [6]. Considering the perfect fit with spin and parity values in the experimental data, it is seen that this statement is correct. This is in excellent agreement with the experimental data. The spin and parity of the first excited state can be identified by some methods such as conversion coefficient, pair creation, lifetime, E/M ratio, angular correlation and nuclear reactions. In this study, the energies of the first $2^+$ excited states of even-even nuclei have been estimated by using the artificial neural network (ANN) method [9]. The data has been borrowed from the Pritychenko et al. [10] in which adopted values cover the Z = 2-104 region including 636 first $2^+$ energy states in the nuclidic chart. The results show that the ANN method is a quite useful method for this type of estimation. Furthermore, the ANN estimated results are compared to the results from nuclear SM [7] calculations for some p, sd, sdpf and pf-shell nuclei given in Table 4 in the reference [10]. According to this comparison, ANN predicts the first $2^+$ energy state energy better than SM calculations. In recent years, ANN has been used in many fields in nuclear physics. It has been used successfully for developing nuclear mass systematic [11, 12], obtaining fission barrier heights [13], obtaining nuclear charge radii [14, 15], estimation of beta decay energies [16] and alpha half-lives calculations of super-heavy nuclei [17]. Since this method is successful in an understanding non-linear relationship between input and output data, layered feed-forward ANN can be used to estimate generate first $2^+$ excited state energy values in even-even nuclei.

## 2. ARTIFICIAL NEURAL NETWORK (ANN)

The ANN method is a strong tool that is used when standard techniques fail [9]. The method mimics the brain functionality of all creatures. Like in the real brain, ANN, which is the base of artificial intelligence, can learn everything by an appropriate algorithm in order to do what it learned. Additionally, artificial intelligence can store lots of data in its memory and keep them in mind through long ages. For this task, ANN is composed of mainly three different layers. The data is taken from outside to the input layer as inputs and the output data is the desired one which is exported from output layers. The number of input neurons depends on the problem and inputs are independent variables. The number of output neurons is the number of output variables which would be estimated. Between these two layers, there is one (or more) additional layer in which data is mainly processed in this layer which is called as hidden layer. In each layer, they have their neurons that are processing units of ANN. Data flows in one direction from input to outputs neurons. Each neuron in the layers is connected to all other neurons in the next layer. Therefore, all neurons in hidden and output layers have at least one own entry. As given in Eq.1 that all these entries ($x_i$) are multiplied by the weight values of their connections ($w_i$) and then summed in order to get the net entries ($n_j$) of the neurons.

$$n_j = \sum_{i=1}^{N} w_i \times x_i \qquad (1)$$

After this step, the neurons are activated by a chosen function and the outcomes of the neurons are transmitted to the neurons in the next layer. In the case of the output neurons, the outcome goes outside.

In the ANN calculations, all data belonging to the given problem has been divided into two main separate sets. The first part of the data (about 80%) is used for the training of ANN in order to get the relationship between input (independent) and the output (dependent) variables. But in order to see the success of the method, it must be tested over another set of data which is the rest (about 20%) of all data. The main task in the training (learning) process is assigning the values to each weighted connection between neurons. In other words, in the training process, it is aimed to find the best weight values which give the best estimation $y_i$ starting from the input $x_i$. Therefore, the weight values are modified until the acceptable deviation level between desired ($d_i$) and neural network ($y_i$) outputs. Generally, the mean square error function (MSE) has been used for the comparison (Eq.2). In order to reach the best weight values, some parameters are tuned up such as hidden layer number, hidden neuron number, learning algorithm, activation function and/or kind of neural network in the training stage. In this study to get the best values, one hidden layer with 12 neurons, Levenberg-Marquardt learning algorithm [18, 19], tangent hyperbolic activation function (Eq.3) and multi-layer feed-forward neural network have been used (Fig.1). The total number of weighted connections is 36 whose 24 of them are from the input to the hidden layer and 12 are from the hidden layer to the output layer.

$$MSE = \frac{1}{N}\sum_{i=1}^{N}(y_i - d_i)^2 \qquad (2)$$

$$G(n_j) = \frac{e^{n_j} - e^{-n_j}}{e^{n_j} + e^{-n_j}} \qquad (3)$$

After a successful training step, the constructed ANN is tested over the training data set which is used in the learning process. By using final weights values, the comparison has been made between neural network outputs and the desired values. However, this result is not sufficient to decide whether the method is good or not. The final weights must also be tested over an unseen data set. The test data set is used for this purpose and if the ANN outputs are also close to the desired values on this data set, it is safely concluded that the ANN is generalized the data. Namely, one can confidently use the constructed ANN with its final weights to solve the problem of the same type of data.

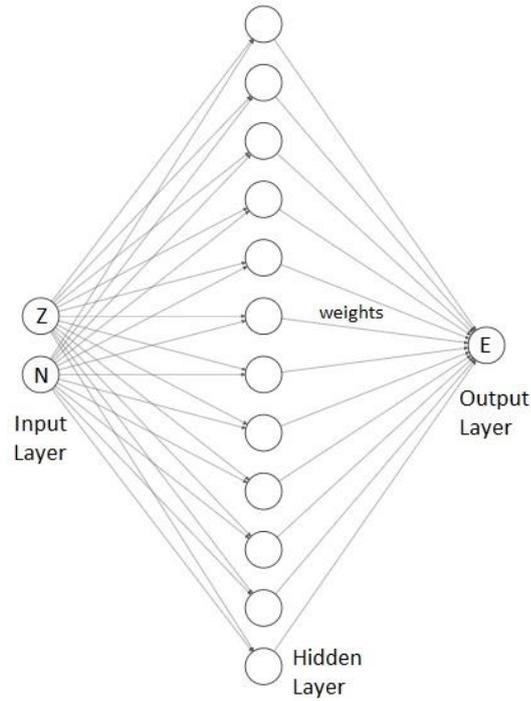

Fig. 1: Used 2-12-1 ANN architecture for the estimation of first $2^+$ excited state energies of even-even nuclei

In the present study, the inputs were proton (Z) and neutron (N) numbers of the atomic nuclei and the output was the first $2^+$ state energy values of the even-even nuclei. Note that the range of activation function is (-1; 1) for the hyperbolic tangent of the hidden layer. Therefore, it can be said that it can potentially be difficult to train cases without normalizing or softening the data. Also generally in the method, the data is normalized or smoothed in order to speed up the learning process and increasing the learning rate. In case of data are always positive and their scales vary drastically, one simple way is to use the logarithm transformation of the data. Thus, we have taken the logarithm of the output values (first $2^+$ state energies) into consideration in the calculations.

## 3. RESULTS AND DISCUSSION

The estimation of the first $2^+$ state energy values in even-even nuclei whose atomic numbers are between 2 and 104 has been performed by using ANN. The total number of nuclei under consideration (which is indicated as data points in the figures) is 631. The adopted literature data (which the recommended values based on all the available experimental data) has been taken from a previous compilation [10]. Due to the different behavior of the regions according to the atomic masses, the estimations have been performed upon two separated parts of the nuclidic chart (Fig.2). there is a dataset of ''adopted'' level and γ -ray transition

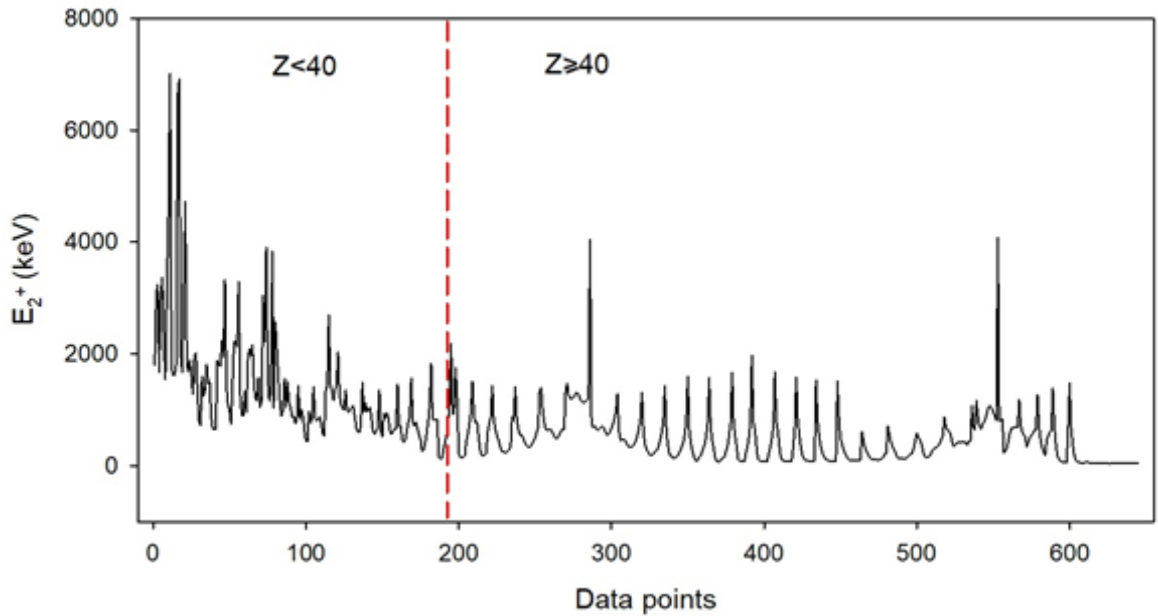

Fig. 2: First $2^+$ excited energy values of nuclei for $Z < 40$ (left) and $Z \geq 40$ (right) regions.

Also, for the heavy elements after atomic number 90, the first excited state energies are lower than the order of 40 keV. The first part includes the nuclei whose atomic mass smaller than 40 and the second one includes greater than or equal 40. As is clear in Fig.2 that the first part includes many more nuclei whose first $2^+$ excited states are higher in energy. The average value of this region is 1393 keV. The average value of the second part is just 479 keV and only for two nuclei, the first 2$^+$ excited states have energy values greater than 2000 keV. These nuclei are $^{132}$Sn and $^{208}$Pb which are doubly magic nuclei and the first 2$^+$ excited state energy values are 4041.20 keV and 4085.52 keV, respectively. In each calculation for different regions, the data in the range has been portioned into two separate parts for training and test of the ANN.

In the estimations on Z < 40 region, 142 data points have been used for the training of the ANN. After applying the ANN method, the results are converted to normal values from their logarithms. The root minimum square error (rmse) and correlation coefficient (r) values have been obtained as 125.1 keV and 0.96, respectively. The correlation coefficient indicates that the method is very useful for the determination of the first excited 2$^+$ state energy values for the atomic nuclei. In Fig.3, we have shown the adopted versus neural network predicted values of even-even nuclei in Z < 40 region. It is clear in the figure that the neural network estimation is in agreement with the adopted values. The data in the figure is concentrated in the (adopted = neural network) line.

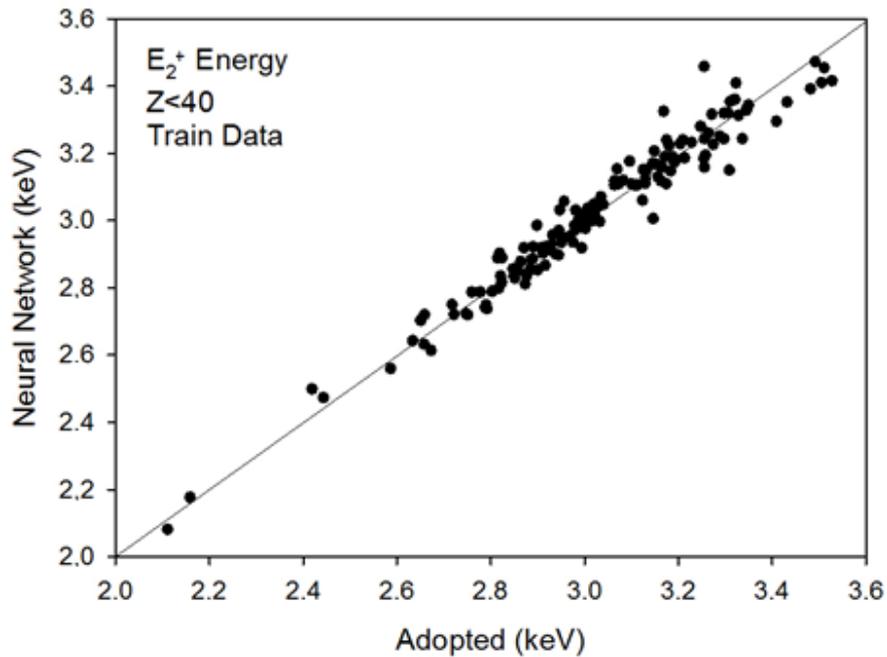

Fig. 3: Adopted [10] vs. ANN estimated first 2$^+$ excited state energies (log) for Z < 40 even-even nuclei in train data

After the training of the ANN, the constructed ANN with final weight values has been used in order to see the generalization ability of the method. In this test stage, the rest of all data (33 data points) has been used for this purpose. The rmse and r values have been obtained as 140.1 keV and 0.94, respectively. The correlation coefficient again indicates that the method is still useful for the determination of the energies for the first excited states in the nuclei. In Fig.4, the values of adopted and neural network predicted first $2^+$ energies of even-even nuclei in Z < 40 region in comparison with each other have been given.

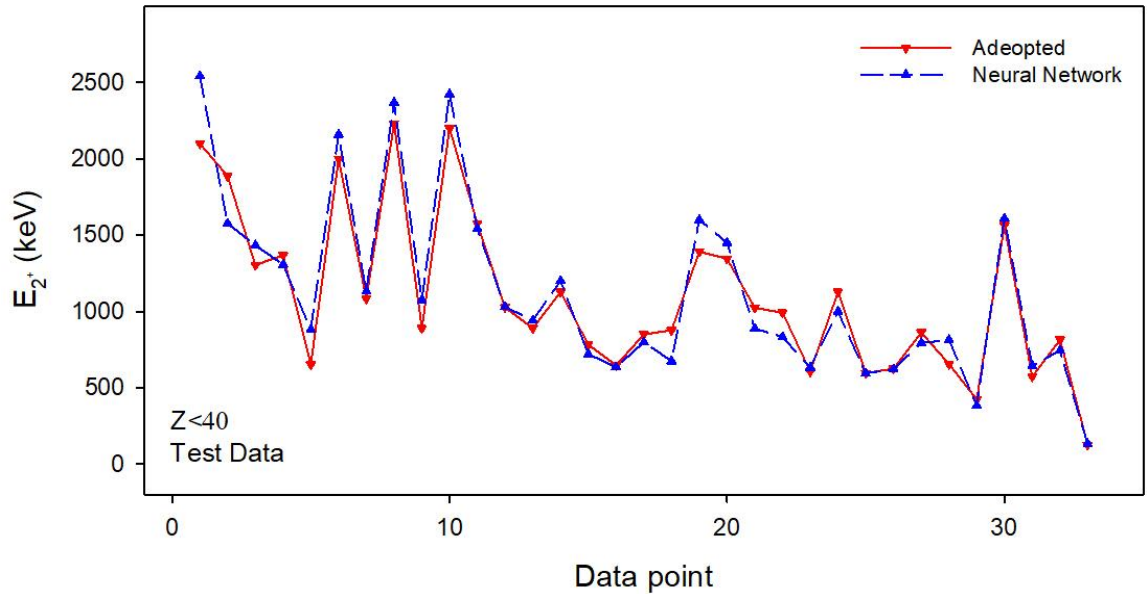

Fig. 4: Adopted [10] and ANN estimated first $2^+$ excited state energies for Z < 40 even-even nuclei in test data

In the second part of the work, we have estimated the first $2^+$ excited state energies in Z ≥ 40 region. 364 data points have been used for the training of the ANN. The results are again presented as real energy values after converting ANN results from logarithmic values. The rmse and r values have been obtained as 82.4 keV and 0.97, respectively. The r has taken its almost maximum value which indicates that the method is still useful for the determination of the first excited $2^+$ state energy values for the atomic nuclei. In Fig.5, we have given the adopted versus neural network predicted energy values of even-even nuclei in $Z \geq 40$ region. The data in the figure is concentrated in the (adopted = neural network) line indicating the neural network estimations are in agreement with the adopted values in the literature.

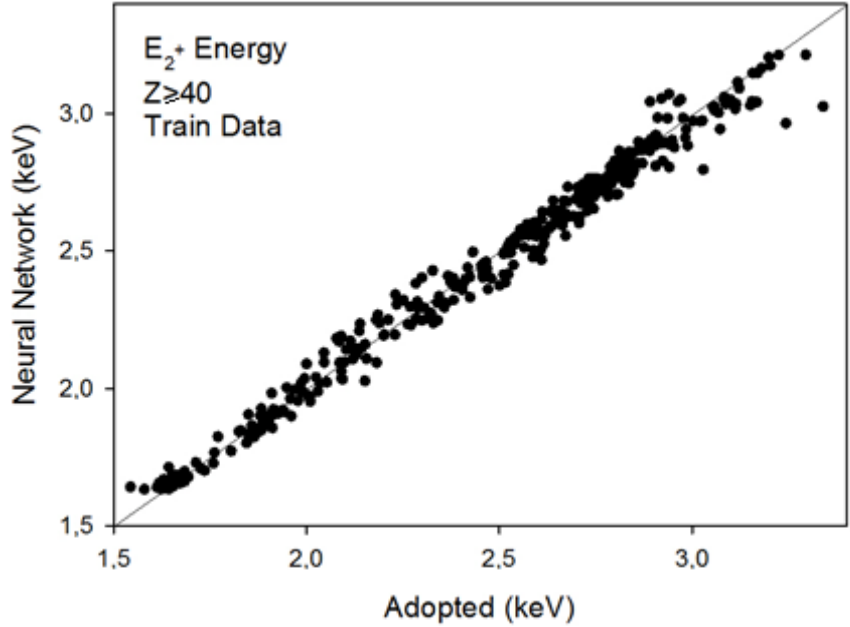

Fig. 5: Adopted [10] vs. ANN estimated first $2^+$ excited state energies (log) for $Z \geq 40$ even-even nuclei in train data

After the training stage for the nuclei in $Z \geq 40$ region, the constructed ANN with final weight values has been tested on the test dataset. The rest of all data (92 data points) has been taken into account for this purpose. The rmse and r values have been obtained as 91.4 keV and 0.94, respectively. The correlation coefficients for both stages indicate that the method is quite useful for the determination of the first excited states of the atomic nuclei. The log values of adopted and neural network predicted energies of even-even nuclei in $Z \geq 40$ region in comparison with each other have been presented in Fig.6. The fact that ANN results do not have the same values as the adopted data indicates that ANN performs well and does not memorize. All rmse and r values in the training and test data for both regions also support this.

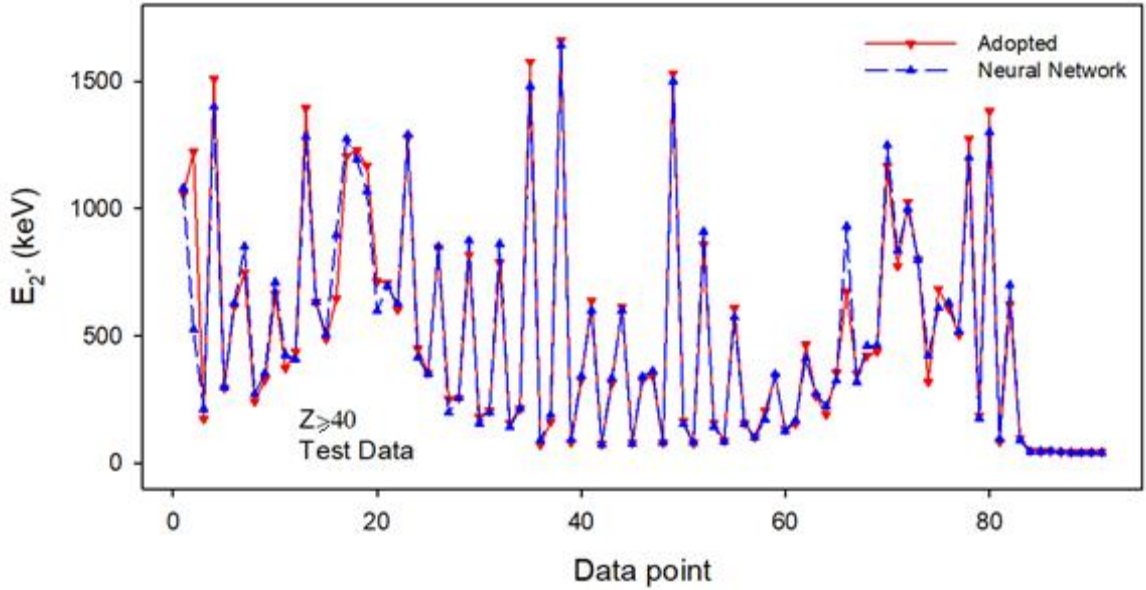

Fig. 6: Adopted [10] and ANN estimated first $2^+$ excited state energies for $Z \geq 40$ even-even nuclei in test data

In the final stage, the ANN estimations for the first $2^+$ excited state energies from the present study and nuclear SM results are both compared to the adopted values of some nuclei given in Table 4 of the reference [10]. SM is one of the most common and best theoretical models for the calculations of nuclear excited states. However, due to some reasons such as the limitation of the model space used in the SM calculations and the not exact values of two-body interaction matrix elements and single-particle energies, the theoretical results cannot be expected to be exactly the same as the experimental ones. As is seen in Table 1 that the mean absolute error values from the adopted values are 136.0 keV and 174.6 keV for ANN estimations and SM results, respectively. For 28 of the total 49 data points, the ANN gives closer results than the SM calculations. Maximum and minimum deviations of ANN estimations from adopted values are 869 keV and 2 keV, respectively. Whereas for SM calculations, these values are 1217 keV and 3 keV. According to the results of ANN and SM that the ANN predictions outperforms SM for 57% of the test nuclei, and leads to an overall reduction of 22% of the test error with 80 parameters of ANN structure. Of course, different results can be obtained by using SM calculations using different effective interactions. However, the present effective interactions used are usually those that work best for the respective nuclei. Therefore, the values given in this reference are considered to be the closest results to the adopted results.

Table 1: Comparison of first 2$^+$ excited state energies from SM calculations [10] and ANN estimations

| A | Z | N | Energy (keV) | | | | |
|---|---|---|---|---|---|---|---|
| | | | Adopted | ANN | SM | ANN Dev. | SM Dev. |
| 6 | 2 | 4 | 1797 | 1801 | 1894 | 4 | 97 |
| 10 | 4 | 6 | 3368 | 3261 | 3704 | 107 | 336 |
| 12 | 4 | 8 | 2102 | 2338 | 3319 | 236 | 1217 |
| 18 | 8 | 10 | 1982 | 2057 | 1999 | 75 | 17 |
| 20 | 8 | 12 | 1674 | 2543 | 1746 | 869 | 72 |
| 22 | 8 | 14 | 3199 | 2768 | 3158 | 431 | 41 |
| 18 | 10 | 8 | 1887 | 1576 | 1999 | 311 | 112 |
| 20 | 10 | 10 | 1634 | 1529 | 1747 | 105 | 113 |
| 22 | 10 | 12 | 1275 | 1636 | 1363 | 361 | 88 |
| 24 | 10 | 14 | 1982 | 1848 | 2111 | 134 | 129 |
| 26 | 10 | 16 | 2018 | 1879 | 2063 | 139 | 45 |
| 28 | 10 | 18 | 1304 | 1433 | 1623 | 129 | 319 |
| 20 | 12 | 8 | 1598 | 1601 | 1746 | 3 | 148 |
| 22 | 12 | 10 | 1247 | 1275 | 1363 | 28 | 116 |
| 24 | 12 | 12 | 1369 | 1306 | 1502 | 63 | 133 |
| 26 | 12 | 14 | 1809 | 1544 | 1897 | 265 | 88 |
| 28 | 12 | 16 | 1474 | 1681 | 1518 | 207 | 44 |
| 30 | 12 | 18 | 1483 | 1434 | 1591 | 49 | 108 |
| 24 | 14 | 10 | 1879 | 1794 | 2111 | 85 | 232 |
| 26 | 14 | 12 | 1796 | 1874 | 1897 | 78 | 101 |
| 28 | 14 | 14 | 1779 | 2159 | 1932 | 380 | 153 |
| 30 | 14 | 16 | 2235 | 2331 | 2266 | 96 | 31 |
| 30 | 16 | 14 | 2211 | 2076 | 2266 | 135 | 55 |
| 32 | 14 | 18 | 1941 | 2084 | 2053 | 143 | 112 |
| 32 | 16 | 16 | 2231 | 2368 | 2160 | 137 | 71 |
| 34 | 16 | 18 | 2128 | 2251 | 2131 | 123 | 3 |
| 36 | 14 | 22 | 1399 | 1168 | 1723 | 231 | 324 |
| 38 | 14 | 24 | 1084 | 1132 | 1395 | 48 | 311 |
| 38 | 16 | 22 | 1292 | 1156 | 1459 | 136 | 167 |
| 40 | 14 | 26 | 986 | 1095 | 1217 | 109 | 231 |
| 40 | 16 | 24 | 904 | 1019 | 942 | 115 | 38 |
| 42 | 16 | 26 | 890 | 1077 | 999 | 187 | 109 |
| 32 | 18 | 14 | 1867 | 1816 | 2053 | 51 | 186 |
| 34 | 18 | 16 | 2091 | 2312 | 2131 | 221 | 40 |
| 58 | 28 | 30 | 1454 | 1485 | 1478 | 31 | 24 |
| 60 | 28 | 32 | 1333 | 1310 | 1474 | 23 | 141 |
| 62 | 28 | 34 | 1173 | 1193 | 1149 | 20 | 24 |
| 66 | 28 | 38 | 1425 | 1647 | 1265 | 222 | 160 |
| 68 | 28 | 40 | 2034 | 1727 | 1963 | 307 | 71 |
| 70 | 28 | 42 | 1260 | 1354 | 1599 | 94 | 339 |
| 72 | 28 | 44 | 1096 | 1036 | 1505 | 60 | 409 |
| 76 | 28 | 48 | 992 | 994 | 1374 | 2 | 382 |
| 62 | 30 | 32 | 954 | 947 | 1013 | 7 | 59 |
| 66 | 30 | 36 | 1039 | 1018 | 950 | 21 | 89 |
| 68 | 30 | 38 | 1077 | 1087 | 879 | 10 | 198 |
| 70 | 30 | 40 | 885 | 909 | 1109 | 24 | 224 |
| 72 | 30 | 42 | 653 | 690 | 1007 | 37 | 354 |
| 76 | 30 | 46 | 599 | 594 | 976 | 5 | 377 |
| 78 | 30 | 48 | 730 | 718 | 1045 | 12 | 315 |
| | | | | | Total Dev. | *136.0* | *174.6* |

## 4. CONCLUSIONS

In this work, the first $2^+$ excited state energies of even-even nuclei in the nuclidic chart have been predicted the first time by using the ANN method. The inputs of the ANN are atomic and neutron numbers of the nuclei. One hidden layer with 12 neurons that gives better results for the problem has been used after several trials. According to estimation performance as obtained from correlation coefficients, the method can be useful for the prediction of the first excited 2+ energy states of nuclei. The method has been applied to the nuclei in two regions of the nuclidic chart. One region contains the nuclei whose atomic number less than 40 and the other includes $Z \geq 40$ region. The rmse values of ANN estimations on the test dataset are 237.9 keV and 91.4 keV for $Z < 40$ and $Z \geq 40$, respectively. The obtained correlation coefficients for $Z < 40$ region are 0.96 and 0.94 for the training and test phase respectively. For $Z \geq 40$ regions, correlation coefficients are almost the same as 0.97 and 0.94 for the training and test phase respectively. Also, the results of the ANN method have been compared to the results from nuclear SM calculations in order to see the success of the method. The ANN results give better results than the theoretical SM calculations. By getting better information about the first 2+ state energy values, more accurate nuclear structure properties might be obtained. Therefore, in order to predict the excited state energy values of the nuclei, the ANN method can be a good alternative with many advantages, such as quick calculation, no need for any complex formulation and easy applicability.

**CONFLICT OF INTEREST**

The authors declare that they have no conflict of interest.